\documentclass[aps,prb,twocolumn,showpacs]{revtex4}
\usepackage{graphicx}
\begin{document}
\title{Radiation-induced magnetoresistance oscillations 
in two-dimensional electron systems under bichromatic irradiation}
\author{X. L. Lei}
\affiliation{Department of Physics, Shanghai Jiaotong University,
1954 Huashan Road, Shanghai 200030, China}

\begin{abstract}

We analyze the magnetoresistance $R_{xx}$ oscillations in high-mobility 
two-dimensional electron systems induced by the combined driving of two radiation fields 
of frequency $\omega_1$ and $\omega_2$,  based on the balance-equation 
approach to magnetotransport for high-carrier-density systems in Faraday geometry.
It is shown that under bichromatic irradiation of $\omega_2\sim 1.5 \omega_1$, 
most of the characterstic peak-valley pairs in the curve of 
$R_{xx}$ versus magnetic field in the case of monochromatic irradiation 
of either $\omega_1$ or $\omega_2$ disappear, except the one around 
$\omega_1/\omega_c\sim 2$ or $\omega_2/\omega_c\sim 3$.
$R_{xx}$ oscillations show up mainly as new peak-valley structures 
around other positions related to multiple photon processes of mixing frequencies
$\omega_1+\omega_2$, $\omega_2-\omega_1$, etc. Many 
minima of these resistance peak-valley pairs can descend down to negative 
with enhancing radiation strength, indicating the possible bichromatic
zero-resistance states. 
      
\end{abstract}


\maketitle

\section{Introduction}

Since the discovery of radiation induced magnetorersistance oscillations (RIMOs) 
and zero-resistance states (ZRS) in ultra-high mobility two-dimensional (2D) electron 
systems,\cite{Zud01,Ye,Mani,Zud03}
tremendous experimental\cite{Yang,Dor03,Mani04,Will,Kovalev,Mani-apl,Zud04,Stud,Dor05,Zud05} 
and theoretical\cite{Ryz,Ryz86,Anderson,Koul,Andreev,Durst,Xie,Lei03,Dmitriev,Ryz05199,
Vav,Mikh,Dietel,Torres,Dmitriev04,Inar,Lei05,Ryz05,Joas05,Ng} efforts have been devoted to study 
this exciting phenomenon and a general understanding of it has been reached.
Under the influence of a microwave radiation of frequency $f=\omega/2\pi$,
the low-temperature magnetoresistance  $R_{xx}$ of a 2D electron gas (EG), 
exhibits periodic oscillation as a function of the inverse magnetic field.
The RIMOs feature the periodical appearance of peak-valley pairs 
around $\omega/\omega_c=1,2,3,4,\cdots$, i.e. 
a maximum at $\omega/\omega_c=j-\delta_j^-$ and 
a minimum at $\omega/\omega_c=j+\delta_j^+$, with $j=1,2,3,4,\cdots$ and 
$0<\delta_j^{\pm}\leq 1/4$.
Here $\omega_c$ is the cyclotron frequency and $\omega/\omega_c=j$ are  
the node points of the oscillation.
With increasing the radiation intensity 
the amplitudes of the peak-valley oscillations increase and the resistance $R_{xx}$ around the 
minima of a few lowest-$j$ pairs can drop down towards negative direction but 
will stop when a vanishing resistance is reached, i.e. ZRS. 

In addition to these basic features, secondary peak-valley pair structures
were also observed experimentally\cite{Mani,Zud03,Dor03,Mani04,Will,Zud04,Dor05} 
and predicted theoretically\cite{Lei03,Lei05} 
around $\omega/\omega_c=1/2,3/2$ and 2/3. They were referred to the effect 
of two- and three-photon processes and their minima were shown also to be able to
develop into negative value when increasing the radiation further.\cite{Lei03}
Recent measurement at 27\,GHz with intensified microwave intensity,\cite{Zud05} 
confirmed these mutliphoton-related peak-valley pairs and the ZRSs
developed from their minima. A careful theoretical analysis with
enhanced radiation convincingly reproduced these structures and predicted more
peak-valley pairs related to multiphoton processes.\cite{Lei06}

Despite intensive experimental and theoretical studies have been done in the case 
of monochromatic irradiation, further investigations beyond this configuration 
are highly desirable for a deeper understanding of this fascinating phenomenon. 
An easy way is to study the response of the 2D system to a bichromatic radiation,
which apparently can not be reduced simply to the superposition of the 
system response to each monochromatic radiation.
Thus the presence of a second radiation of different frequency 
could provide additional insight into the problem of microwave-driven 2D electron system.

Theoretically, effect of a bichromatic irradiation on transport of a 2D
electron gas was investigated within a model of a clean classical gas, 
in which Kohn's theorem is violated entirely due to nonparabolicity
without invoking Landau quantization.\cite{Joas04} Some  
features that are specific to the bichromatic case, such as
new domain of magnetic field within which the diagonal conductivity 
is negative, were revealed.

Experimentally, Zudov {\it et al.}\cite{Zud05} recently measured the magnetoresistance of
a high-mobility 2D electron system under the combined driving of two radiation fields 
of frequency $\omega_1$ and $\omega_2$, and disclosed features of RIMOs which are quite 
different from those under monochromatic radiation of frequency $\omega_1$
or $\omega_2$. They detected a new resistance
minimum under bichromatic microwave radiation, which seems to originate 
from a frequency mixing process, possibly a precursor 
of bichromatic ZRS. 

In this paper we report our studies on microwave photoresistance 
response of high-mobility two-dimensional electron systems  
under bichromatic irradiation, based on 
the balance-equation approach to magnetotransport for 
high-carrier-density systems, extended to the case of 
simultaneous driving of two radiation fields 
of different frequencies. The balance-equation approach, though 
semiclassical in nature, has been shown to capture the essence of
this radiation-induced nonlinear magnetotransport quantitatively.\cite{Lei03,Lei05}
Under bichromatic irradiation of $\omega_2\sim 1.5 \omega_1$, we find that
most of the characteristic peak-valley pairs in the curve of 
$R_{xx}$ versus magnetic field in th case of monochromatic irradiation 
of either $\omega_1$ or $\omega_2$ disappear, except the one around 
$\omega_1/\omega_c\sim 2$ or $\omega_2/\omega_c\sim 3$.
$R_{xx}$ oscillations show up mainly as new peak-valley structures 
around other positions related to multiple photon processes of mixing frequencies
$\omega_1+\omega_2$, $\omega_2-\omega_1$, etc.

\section{Balance equations under bichromatic radiation}

The derivation of balance equations under bichromatic radiation follows
that of monochromatic radiation.\cite{Lei05} The nature 
of the balance-equation approach and its applicability to radiation-driven
magnetotransport in high-mobilty 2D electron systems of high carrier density,
was examined closely in Ref.\,\onlinecite{Lei05}. We refer the readers    
to it for detail.

In the case of bichromatic radiation we consider that a dc or slowly time-varying 
electric field ${\bf E}_0$ and two high frequency (HF) fields 
\begin{equation}
{\bf E}_1(t)\equiv{\bf E}_{1s} \sin(\omega_1 t)+{\bf E}_{1c}\cos(\omega_1 t)
\end{equation}
and
\begin{equation}
{\bf E}_2(t)\equiv{\bf E}_{2s} \sin(\omega_2 t)+{\bf E}_{2c}\cos(\omega_2 t)
\end{equation} 
are applied simultaneously in a quasi-2D system consisting of $N_{\rm e}$ 
interacting electrons in a unit area of the $x$-$y$ plane, 
together with a magnetic field ${\bf B}=(0,0,B)$ along the $z$ direction.
The frequencies $\omega_1$ and $\omega_2$ are high enough and their difference
is large enough that $\omega_1$ and $\omega_2$, as well as $|n_1\omega_1-n_2\omega_2|$
(for arbitrary integers $n_1$ and $n_2$), 
are all much larger than $1/\tau_0$, where $\tau_0$ stands for the scale of the time 
variation of slowly varying field ${\bf E}_0$, or the time scale within which 
one carries out the transport measurement, whichever the shorter.  
The approach is based on the separation of the center-of-mass motion 
from the relative electron motion of the electrons and 
describes the transport state of a high-carrier-density many-electron system
under radiation fields in terms of a rapidly time-varying electron drift velocity
oscillating at both base radiation frequencies, 
$
{\bf v}(t)={\bf v}_1(t)+{\bf v}_2(t)$, with 
\begin{equation}
{\bf v}_1(t)={\bf v}_{1c} \cos(\omega_1 t)+{\bf v}_{1s} \sin(\omega_1 t),
\end{equation}
\begin{equation}
{\bf v}_2(t)=
{\bf v}_{2c} \cos(\omega_2 t)+{\bf v}_{2s} \sin(\omega_2 t),
\end{equation}
together with another part ${\bf v}_0$ for the slowly varying 
electron drift motion, as well as an electron temperature $T_{\rm e}$ 
characterizing the electron heating.\cite{Lei85,Liu} 
In the case of ultra-clean electron gas at low temperatures,
the slowly time-varying quantities ${\bf v}_0$ and $T_{\rm e}$ satisfy 
the following force- and energy-balance equations:\cite{Lei03,Lei05}
\begin{equation}
m\frac{d{\bf v}_{0}}{dt}=e{\bf E}_{0}+ e ({\bf v}_0 \times {\bf B})+
\frac{{\bf F}_0}{N_{\rm e}},\label{eqv0}
\end{equation}
\begin{equation}
N_{\rm e}{\bf E}_0\cdot {\bf v}_0+S_{\rm p}- W=0,
\label{eqne}
\end{equation}
with ${\bf v}_{1c}$ and ${\bf v}_{1s}$ determined by
\begin{eqnarray}
-{m\omega_1}{\bf v}_{1c}&=&{e{\bf E}_{1s}}
+e({\bf v}_{1s}\times
{\bf B}),\label{eqv1}\\
{m\omega_1}{\bf v}_{1s}&=&{e{\bf E}_{1c}}
+e({\bf v}_{1c}
\times {\bf B});\label{eqv2}
\end{eqnarray}
and 
${\bf v}_{2c}$ and ${\bf v}_{2s}$ determined by
\begin{eqnarray}
-{m\omega_2}{\bf v}_{2c}&=&{e{\bf E}_{2s}}
+e({\bf v}_{2s}\times
{\bf B}),\label{eqv3}\\
{m\omega_2}{\bf v}_{2s}&=&{e{\bf E}_{2c}}
+e({\bf v}_{2c}
\times {\bf B}).\label{eqv4}
\end{eqnarray}
Here $e$ and $m$ are the electron charge and effective mass, 
\begin{eqnarray}
&&\hspace*{-0.6cm}{\bf F}_{0}=\sum_{{\bf q}_\|}\left| U({\bf q}_\|)\right| ^{2}
\sum_{n_1,n_2=-\infty }^{\infty }
{J}_{n_1}^{2}(\xi_1){J}_{n_2}^{2}(\xi_2)\nonumber\\
&&\hspace*{2cm}\times\,{\bf q}_\|\,\Pi _{2}({\bf q}_\|,\omega_0-n_1\omega_1-n_2\omega_2) \label{eqf0}
\end{eqnarray}
is the damping force of the moving center-of-mass,
\begin{eqnarray}
&&\hspace*{-0.9cm}S_{\rm p}=\sum_{{\bf q}_\|}\left| U({\bf q}_\|)\right| ^{2}
\sum_{n_1,n_2=-\infty }^{\infty }
{J}_{n_1}^{2}(\xi_1){J}_{n_2}^{2}(\xi_2)\nonumber\\
&&\hspace*{0.3cm}\times\,(n_1\omega_1+n_2\omega_2)\,
\Pi _{2}({\bf q}_\|,\omega_0-n_1\omega_1-n_2\omega_2) \label{eqsp}
\end{eqnarray} 
is the averaged rate of the electron energy absorption from the HF fields, and
\begin{eqnarray}
&&\hspace*{-0.6cm}W=\sum_{{\bf q}}\left| M({\bf q})\right| ^{2}
\sum_{n_1,n_2=-\infty }^{\infty }
{J}_{n_1}^{2}(\xi_1){J}_{n_2}^{2}(\xi_2)\nonumber\\
&&\hspace*{1.0cm}\times\,\Omega_{\bf q}\,
\Lambda _{2}({\bf q},\omega_0+\Omega _{{\bf q}}-n_1\omega_1-n_2\omega_2) 
\label{eqw}
\end{eqnarray}
is the average rate of the electron energy dissipation to the lattice. 
In the above equations, $J_{n_1}(\xi_1)$ and $J_{n_2}(\xi_2)$
are Bessel functions of order $n_1$ and $n_2$,
$
\xi_1\equiv \sqrt{({\bf q}_\|\cdot {\bf v}_{1c})^2+
({\bf q}_\|\cdot {\bf v}_{1s})^2}/{\omega_1}
$,
$
\xi_2\equiv \sqrt{({\bf q}_\|\cdot {\bf v}_{2c})^2+
({\bf q}_\|\cdot {\bf v}_{2s})^2}/{\omega_2}
$, and 
$\omega_0\equiv {\bf q}_\|\cdot {\bf v}_0$.
Here $q_{\|}\equiv (q_x,q_y)$ stands for the in-plane wavevector,
$U({\bf q}_\|)$ is the effective impurity scattering potential, 
and $\Pi_2({\bf q}_\|,\Omega)$ is the imaginary part of the electron density
correlation function of the quasi-2D system in the magnetic field.
In Eq.\,(\ref{eqw}), ${\bf q}$ represents the 3D wavevector $({\bf q}_{\|},q_z)$ 
plus the branch index $\lambda$, and the summation is for
all possible 3D phonon modes of frequency $\Omega_{\bf q}$ having electron-phonon
scattering matrix element $M({\bf q})$.
$\Lambda_2({\bf q},\Omega)=2\Pi_2({\bf q}_\|,\Omega)
[n(\Omega_{\bf q}/T)-n(\Omega/T_{\rm e})]
$ (with $n(x)\equiv 1/({\rm e}^x-1)$)
is the imaginary part of the electron-phonon correlation function.
The $\Pi_2({\bf q}_{\|}, \Omega)$ function of a 2D
system in a magnetic field can be expressed 
in the Landau representation:\cite{Ting}
\begin{eqnarray}
&&\hspace{-0.7cm}\Pi _2({\bf q}_{\|},\Omega ) =  \frac 1{2\pi
l_{\rm B}^2}\sum_{n,n'}C_{n,n'}(l_{\rm B}^2q_{\|}^2/2) 
\Pi _2(n,n',\Omega),
\label{pi_2}\\
&&\hspace{-0.7cm}\Pi _2(n,n',\Omega)=-\frac2\pi \int d\varepsilon
\left [ f(\varepsilon )- f(\varepsilon +\Omega)\right ]\nonumber\\
&&\,\hspace{2cm}\times\,\,{\rm Im}G_n(\varepsilon +\Omega){\rm Im}G_{n'}(\varepsilon ),
\end{eqnarray}
where
$
C_{n,n+l}(Y)\equiv n![(n+l)!]^{-1}Y^le^{-Y}[L_n^l(Y)]^2
$
with $L_n^l(Y)$ the associate Laguerre polynomial, 
$l_{\rm B}\equiv\sqrt{1/|eB|}$ is the magnetic length, 
$f(\varepsilon )=\{\exp [(\varepsilon -\mu)/T_{\rm e}]+1\}^{-1}$ 
is the Fermi distribution function at electron temperature $T_{\rm e}$. 
The density of states of the $n$-th Landau level
is modeled with a Gaussian form:\cite{Ando}
\begin{equation}
{\rm Im}G_n(\varepsilon)=-(\sqrt{2\pi}/\Gamma)
\exp[-{2(\varepsilon-\varepsilon_n)^2}/{\Gamma^2}], \label{gauss}
\end{equation}
having a half-width
\begin{equation}
\Gamma=\left(\frac{8e\omega_c\alpha}{\pi m \mu_0}\right)^{1/2}
\end{equation}
around the level center $\varepsilon_n=n\omega_c$. Here $\omega_c=eB/m$
is the cyclotron frequency, 
$\mu_0$ is the linear mobility at lattice temperature $T$ 
in the absence of the magnetic field, 
and $\alpha$ is a semi-empirical 
parameter to take account of the difference of 
the transport scattering time $\tau_m$ determining the mobility $\mu_0$,    
from the single particle lifetime $\tau_s$ related to Landau level broadening.

For time-independent ${\bf v}_0$, we immediately deduce 
the transverse and longitudinal dc resistivities from Eq.\,(\ref{eqv0}):
\begin{equation}
R_{xy}=B/N_{\rm e}e,
\end{equation} 
\begin{equation}
R_{xx}=-{\bf F}_0\cdot{\bf v}_0/({N_{\rm e}^2e^2v_0^2}). \label{rxxg}
\end{equation}
The (linear) magnetoresistivity is its ${\bf v}_0\rightarrow 0$ limit: 
\begin{eqnarray}
&&\hspace*{-1cm}R_{xx}=-\frac{1}{N_{\rm e}^2 e^2}\sum_{{\bf q}_\|} q_x^2|
U({\bf q}_\|)|^2\times\nonumber\\
&&\sum_{n_1,n_2=-\infty }^\infty  
J_{n_1}^2(\xi_1)J_{n_2}^2(\xi_2)
\left. 
\frac {\partial \Pi_2}{\partial\, \Omega }\right|_{\Omega =n_1\omega_1+n_2\omega_2}.
\label{rxx}
\end{eqnarray}

We assume that the 2DEG is contained in a thin sample suspended in vacuum 
at plane $z=0$.  
When both electromagnetic waves illuminate upon the plane perpendicularly 
with the incident electric fields 
${\bf E}_{\rm i1}(t)={\bf E}_{{\rm i1}s}\sin(\omega_1 t)+ {\bf E}_{{\rm i1}c}\cos(\omega_1 t)$ 
and
${\bf E}_{\rm i2}(t)={\bf E}_{{\rm i2}s}\sin(\omega_2 t)+ {\bf E}_{{\rm i2}c}\cos(\omega_2 t)$, 
the HF electric fields in the 2DEG, determined by the electrodynamic equations, 
are   
\begin{eqnarray}
&&{\bf E}_1(t)=\frac{N_{\rm e}e\,{\bf v}_1(t)}{2\epsilon_0 c}+{\bf E}_{\rm i1}(t),\\
&&{\bf E}_2(t)=\frac{N_{\rm e}e\,{\bf v}_2(t)}{2\epsilon_0 c}+{\bf E}_{\rm i2}(t).
\end{eqnarray}
Using this ${\bf E}_1(t)$ in Eqs.\,(\ref{eqv1}) and (\ref{eqv2}), and ${\bf E}_2(t)$
in Eqs.\,(\ref{eqv3}) and (\ref{eqv4}), the oscillating velocity
${\bf v}_{1c}$ and ${\bf v}_{1s}$ (and thus the argument $\xi_1$) 
are explicitly expressed in terms of incident
field ${\bf E}_{{\rm i1}s}$ and ${\bf E}_{{\rm i1}c}$, 
and the oscillating velocity ${\bf v}_{2c}$ and ${\bf v}_{2s}$ (and thus the argument $\xi_2$)
are explicitly expressed in terms of incident
field ${\bf E}_{{\rm i2}s}$ and ${\bf E}_{{\rm i2}c}$.
Therefore, in the case of weak measuring current limit ($v_0\rightarrow 0$) we
need only to solve the energy balance equation $S_{\rm p}-W=0$ to obtain the 
electron temperature $T_{\rm e}$ under given incident radiation fields, before
directly calculating the linear magnetoresistivity from Eq.\,(\ref{rxx}). 

Within certain field range, the magnetoresistivity $R_{xx}$ given by Eq.\,(\ref{rxxg})
can be negative at small $v_0$, but will change towards the positive direction with increasing $v_0$ 
and passes through zero at a finite $v_0$,\cite{Lei03} implying that 
the time-independent small-current solution is unstable and a spatially nonuniform\cite{Andreev} or
a time-dependent solution\cite{Ng} may develop, which exhibits measured zero resistance.
Therefore we identify the region where a negative dissipative magnetoresistance
develops as that of the ZRS.

The summations over $n_1$ and $n_2$ 
in Eqs.\,(\ref{eqf0}), (\ref{eqsp}) and (\ref{eqw}) include all possible
electron transition processes assisted by real and virtual photons of frequency
$\omega_1$ and $\omega_2$.
The $n_1>0$ terms represent all possible electron transitions with simultaneous absorption of
$n_1$ photons of frequency $\omega_1$, while $n_1<0$ terms represent all possible electron transitions 
with simultaneous emission of $|n_1|$ photons of frequency $\omega_1$. We call these
electron transitions the $|n_1|$ $\omega_1$-photon assisted processes. 
The same for $|n_2|$ $\omega_2$-photon assisted processes.
The $n_1=0$ terms represent electron transitions assisted by virtual $\omega_1$-photons
(all possible emission and absorption of same number of $\omega_1$-photons),
and $n_2=0$ terms represent electron transitions assisted by virtual $\omega_2$-photons.

\section{Numerical results}

 As indicated by experiments,\cite{Umansky} in ultra-clean GaAs-based 
2D samples having mobility of order of $10^{3}$\,m$^{2}$/Vs, the remote donor 
scattering is responsible for merely $\sim 10\%$ or less of the total momentum 
scattering rate. The dominant contribution to the momentum scattering comes 
from short-range scatterers such as residual impurities or defects in the background.
Therefore, as in Ref.\,\onlinecite{Lei05}, 
we assume that the dominant elastic scatterings contributing to the resistance and energy absorption
are due to short-range impurities 
randomly distributed throughout the GaAs region in the numerical calculations.  
The impurity densities are determined by the 
requirement that electron total linear mobility at zero magnetic field equals 
the giving value at lattice temperature $T$.
Since RIMO measurements are at low temperatures, the direct
phonon contributions to $S_{\rm p}$ and ${\bf F}_0$ ($R_{xx}$) can be neglected.
Nevertheless, to calculate the electron energy dissipation to the lattice, $W$,
we take account of scatterings from bulk longitudinal acoustic (LA) and transverse acoustic (TA) 
phonons (via the deformation potential and piezoelectric couplings),
as well as from longitudinal optical (LO) phonons (via the Fr\"{o}hlich coupling) 
in the GaAs-based system. 
The relevant material and coupling parameters 
are taken typical values of GaAs:\cite{Lei851}
electron effective mass $m=0.068\,m_{\rm e}$ ($m_{\rm e}$ is the
free electron mass), transverse sound speed $v_{\rm st}=2.48\times 10^3$\,m/s,
longitudinal sound speed $v_{\rm sl}=5.29\times 10^3$\,m/s, acoustic 
deformation potential $\Xi=8.5$\,eV, piezoelectric constant $e_{14}=
1.41\times 10^9$\,V/m, dielectric constant $\kappa=12.9$, 
material mass density $d=5.31$\,g/cm$^3$.  

The numerical calculations are performed for $x$-direction 
(parallel to ${\bf E}_0$) linearly polarized incident microwave fields  
[${\bf E}_{{\rm i1}s}=(E_{{\rm i1}},0), {\bf E}_{{\rm i1}c}=0$ 
and ${\bf E}_{{\rm i2}s}=(E_{{\rm i2}},0), {\bf E}_{{\rm i2}c}=0$]. 

\begin{figure}
\includegraphics [width=0.45\textwidth,clip=on] {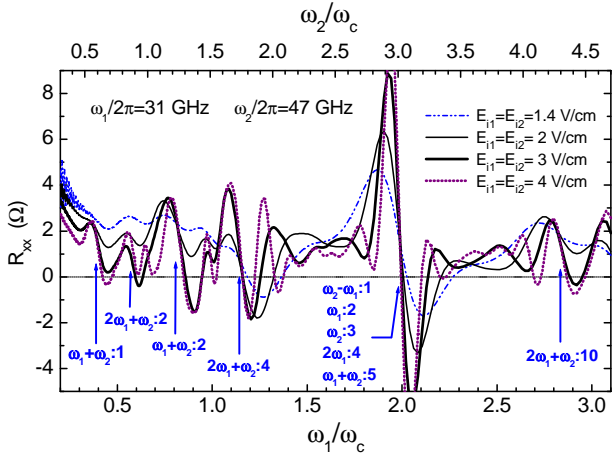}
\vspace*{-0.2cm}
\caption{(Color online) The magnetoresistivity $R_{xx}$ versus the inverse magnetic field $1/B$
for a GaAs-based 2DEG 
with $N_{\rm e}=3.0\times 10^{15}$\,m$^{-2}$, $\mu_0=2000$\,m$^2$/Vs and $\alpha=5$, 
irradiated simultaneously by to two microwaves with frequencies $\omega_1/2\pi=31$\,GHz
and $\omega_2/2\pi=47$\,GHz having four sets of incident amplitudes 
$E_{{\rm i1}}=E_{{\rm i2}}=1.4,2,3,$ and 4\,V/cm at lattice temperature $T=1$\,K. 
$\omega_c=eB/m$ is the cyclotron frequency at the magnetic field $B$.}
\label{fig1}
\end{figure}
\begin{figure}
\includegraphics [width=0.45\textwidth,clip=on] {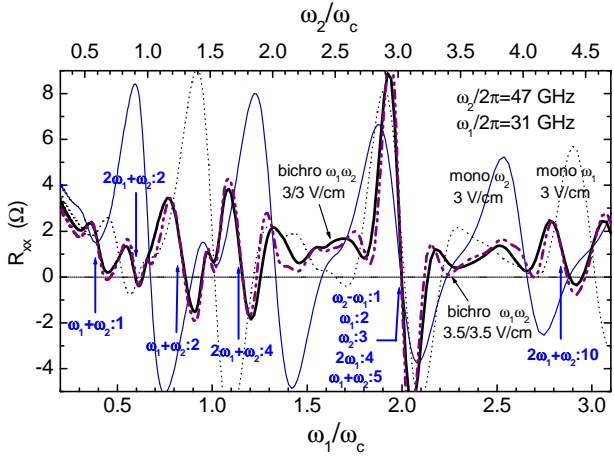}
\vspace*{-0.2cm}
\caption{(Color online) The magnetoresistivity $R_{xx}$ versus the inverse magnetic field $1/B$
for the same system as described in Fig.\,1. The thin curves are under monochromatic
irradiation either of frequency $\omega_1/2\pi=31$\,GHz and incident amplitude
$E_{{\rm i}}=3$\,V/cm (dot curve), or of frequency $\omega_1/2\pi=47$\,GHz and 
incident amplitude $E_{{\rm i}}=3$\,V/cm (solid curve).
The thick curves are under bichromatic irradiation of 
frequencies $\omega_1/2\pi=31$\,GHz and $\omega_2/2\pi=47$\,GHz having 
incident amplitudes $E_{{\rm i1}}=E_{{\rm i2}}=3$\,V/cm (solid curve) or 
$3.5$\,V/cm (dash-dot curve). The lattice temperature is $T=1$\,K.}
\label{fig2}
\end{figure}
\begin{figure}
\includegraphics [width=0.45\textwidth,clip=on] {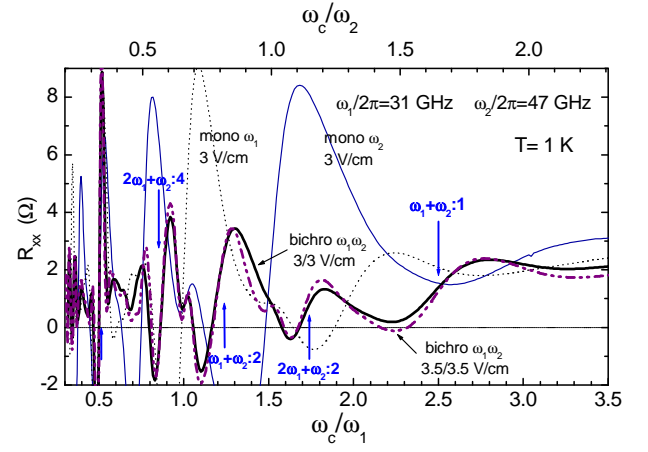}
\vspace*{-0.2cm}
\caption{(Color online) The same as Fig.\,2, but magnetoresistivity $R_{xx}$ is shown
as a function of the magnetic field $B$.}
\label{fig3}
\end{figure}

Figure 1 shows the calculated magnetoresistivity $R_{xx}$ 
versus the inverse magnetic field for a GaAs-based 2D system having electron density 
$N_{\rm e}=3.0\times 10^{15}$\,m$^{-2}$, 
linear mobility $\mu_0=2000$\,m$^2$/Vs and broadening parameter 
$\alpha=5$, simultaneously
irradiated by two microwaves of frequencies $\omega_1/2\pi=31$\,GHz
and $\omega_2/2\pi=47$\,GHz with four sets of incident amplitudes 
$E_{{\rm i1}}=E_{{\rm i2}}=1.4,2,3,$ and 4\,V/cm at lattice temperature $T=1$\,K.
We see that the bichromatic photoresponse of $R_{xx}$ is unlike that of monochromatic
irradiation. Most of the characterizing peak-valley pairs in the curve of 
$R_{xx}$ versus magnetic field subjected to monochromatic irradiation 
of either $\omega_1$ or $\omega_2$ disappear. The only exception is the 
peak-valley pair around $\omega_1/\omega_c\sim 2$ or $\omega_2/\omega_c\sim 3$,
which, on the contrary, is somewhat enhanced in bichromatic irradiation. 
New peak-valley pairs appear. Their amplitudes generally increase with increasing 
the incident field strengths as indicated in the figure. Most of these pairs exhibit
essentially fixed node positions while growing amplitudes and their minima can drop down 
into negative. 
 
As in the case of monochromatic illumination, the appearance of magnetoresistance 
oscillation in the bichromatic irradiation comes from real photon-assisted 
electron transitions between different Landau levels as indicated in the summation of
the electron density-correlation function in Eq.\,(\ref{pi_2}). 
Apparently, all multiple photon processes related to $\omega_1$ monochromatic radiation
and $\omega_2$ monochromatic radiation are included. In addition, there are 
multiple photon processes related to mixing $\omega_1$ and $\omega_2$ radiation.

With three positive integers $n_1,n_2$ and $l$ to characterize a multiple photon assisted electron
transition, we use the symbol $n_1\omega_1+n_2\omega_2:l$ to denote a process during which 
an electron jumps across $l$ Landau level spacings with simultaneous absorption of  
$n_1$ photons of frequency $\omega_1$ and $n_2$ photons of frequency $\omega_2$,
or simultaneous emission of  
$n_1$ photons of frequency $\omega_1$ and $n_2$ photons of frequency $\omega_2$; 
and use the symbol
$n_1\omega_1-n_2\omega_2:l$ to denote a process during which 
an electron jumps across $l$ Landau level spacings with simultaneous absorption of  
$n_1$ photons of frequency $\omega_1$ and emission of $n_2$ photons of frequency 
$\omega_2$,
or simultaneous emission of $n_1$ photons of frequency $\omega_1$ 
and absorption of $n_2$ photons of frequency $\omega_2$ 
(thus the symbol $n_2\omega_2-n_1\omega_1:l$ has the same meaning as 
$n_1\omega_1-n_2\omega_2:l$).
The symbol $n_1\omega_1:l$ indicates a process during which 
an electron jumps across $l$ Landau level spacings with the assistance of 
$n_1$ real (emission or absorption) 
$\omega_1$-photons and virtual $\omega_2$-photons. So does the symbol $n_2\omega_2:l$.  
We find that the processes represented by $n_1\omega_1+n_2\omega_2:l$   
contribute, in the $R_{xx}$-versus-$\omega_c/\omega$ curve,
a structure consisting of a minimum and a maximum on both sides of 
$\omega_c/\omega_1=n_1(1+n_2\omega_2/n_1\omega_1)/l$
or $\omega_c/\omega_2=n_2(1+n_1\omega_1/n_2\omega_2)/l$.
And those by $n_1\omega_1-n_2\omega_2:l$ (assume $n_1\omega_1>n_2\omega_2$)
contribute a minimum-maximum pair around 
$\omega_c/\omega_1=n_1(1-n_2\omega_2/n_1\omega_1)/l$
or $\omega_c/\omega_2=n_2(n_1\omega_1/n_2\omega_2-1)/l$.

From the structure of $R_{xx}$ curves with increasing radiation 
strength in Fig.\,1 we can clearly identify several peak-valley pairs
characteristic of the bichromatic irradiation.
The strongest peak-valley pair is around $\omega_1/\omega_c\sim 2$ or $\omega_2/\omega_c\sim 3$,
which can be referred to the joint contribution from single and multiple photon
processes with mono-$\omega_1$, mono-$\omega_2$ and 
mixing $\omega_1$ and $\omega_2$, such as $\omega_2-\omega_:1$, $\omega_1:2$, $\omega_2:3$, $2\omega_1:4$,
$\omega_1+\omega_2:5$, $\cdots$ . The other peak-valley pairs which are clearly 
identified include that around 
$\omega_1/\omega_c\sim 0.4$ referred to $\omega_1+\omega_2:1$, $\cdots$ ; that 
around $\omega_1/\omega_c\sim 0.57$ referred to $2\omega_1+\omega_2:2$, $\cdots$ ; that
around $\omega_1/\omega_c\sim 0.8$ referred to $\omega_1+\omega_2:2$, $\cdots$ ; that
around $\omega_1/\omega_c\sim 1.14$ referred to $2\omega_1+\omega_2:4$, $\cdots$ ; that
around $\omega_1/\omega_c\sim 2.84$ referred to $2\omega_1+\omega_2:10$, $\cdots$ .
They are indicated in the figure. The minima of all these peak-valley pairs can drop
down to negative when increasing the strengths of the radiation field, indicating the 
the possible locations of the bichromatic ZRSs.

To compare the photoresponse under bichromatic radiation with 
those under monochromatic radiation we plot in Fig.\,2 the 
magnetoresistivity $R_{xx}$ versus the inverse magnetic field
for the above system under simultaneous irradiation of two microwaves 
having frequencies $\omega_1/2\pi=31$\,GHz and $\omega_2/2\pi=47$\,GHz 
with incident amplitudes $E_{{\rm i1}}=E_{{\rm i2}}=3$ and 3.5\,V/cm,
together with those subjected to single microwave radiation of frequency 
$\omega_1/2\pi=31$\,GHz with incident amplitude $E_{{\rm i}}=3$\,V/cm, 
or of frequency $\omega_1/2\pi=31$\,GHz with incident amplitude 
$E_{{\rm i}}=3$\,V/cm.

This figure is redrawn in Fig.\,3 showing $R_{xx}$ as a function of the magnetic field
to give a clearer view in the higher magnetic field side. It indicates 
a possible bichromatic ZRS around $\omega_c/\omega_1\sim 2.2$ or 
$\omega_c/\omega_2\sim 1.5$ arising from $R_{xx}$ droping to negative 
at the minimum of the valley-peak pair around $\omega_c/\omega_1\sim 2.5$
associated with the mixing biphoton process $\omega_1+\omega_2:1$.
This result is in agreement with the recent experimental observation
on the possible precursor of bichromatic ZRS.\cite{Zud05}

{\it Note added}: After the acceptance of this paper by Phys. Rev. B, we read 
a further report of the bichromatic microwave photoresistance measurement.\cite{Zud06}
The experimental peak/valley positions exhibit a good agreement 
with the present theoretical results in a wide magnetic field range.
Our predicted peaks around $\omega_c/\omega_1=0.77, 0.52, 0.37$,
and valley at $\omega_c/\omega_1=0.33$ (Fig.\,3), quite accurately 
($m=0.068\,m_{\rm e}$)
reproduce the observed peaks around $B\approx 570, 380, 275$\,G 
and the valley around $B\approx 250$\,G. 

This work was supported by Projects of the National Science Foundation of China
and the Shanghai Municipal Commission of Science and Technology.

\end{document}